\documentclass{kapproc} 

\usepackage{graphicx}

\upperandlowercase

\nochapequationnumber 

\normallatexbib 

\begin{document}

\articletitle{Black holes: physics and astrophysics}

\articlesubtitle{Stellar-mass, supermassive and primordial black holes}

\author{Jacob D. Bekenstein}
\affil{Racah Institute of Physics, Hebrew University of
Jerusalem, Givat Ram, Jerusalem 91904, Israel\\
Jefferson Physical Laboratory, Harvard University, Cambridge, MA 02138
\email{bekenste@vms.huji.ac.il}}

\begin{abstract}
I present an elementary primer of black hole physics, including its general relativity basis, all peppered with astrophysical illustrations.  Following a brief review of the
process stellar collapse to a black hole, I discuss the gravitational redshift, particle trajectories in gravitational fields, the Schwarzschild and Kerr solutions to Einstein's equations, orbits in Schwarzschild and in Kerr geometry, and the dragging of inertial frames.  I follow with a brief review of galactic X-ray binary systems with known black holes, stressing the QPO phenomenon in particular.

I then discuss the evidence from AGN for the existence of supermassive black holes in galaxy nuclei, as well as evidence for such black holes in ordinary galaxy nuclei.  I use the free motion of gas parcels to illustrate aspects of accretion disks around black holes, showing how to calculate energy efficiency and surface emissivity of disks, and the rate of black hole spin-up.  I recall the primary methods for black hole mass determination, the correlation of black hole mass with the stellar velocity dispersion of its neighborhood, and implications for the origin of supermassive black holes.  Finally, I consider the formation of primordial black holes, and calculation of their mass spectrum at present in the case of scale invariant primordial inhomogeneities.

\end{abstract}


\subsection{Introduction}
\label{sec:intro1} 

From the Newtonian standpoint  a black hole (BH) is a compact
object whose surface gravitational potential approaches the
square of the speed of light.  Thus in contrast with other
astrophysical objects, the radius  $R$ of a BH scales
linearly with its mass $M$:
\begin{equation} 
R= GM/c^2\approx 1.47
(M/M_\odot)\ {\rm km}.
\end{equation} 
General Relativity (GR) {\it requires\/} the existence of BHs.  Not only radius but several other BH properties scale simply with mass.  For example the average mass density $\bar\rho\equiv 3M/4\pi R^3$ scales as $M^{-2}$,
\begin{equation}
\bar\rho\approx 1.5\times 10^{20}(M_\odot/M)^2\, {\rm Kg\, m}^{-3},
\label{density}
\end{equation}
so that massive BHs are tenuous and light BHs are dense.  Classical BH physics lacks a special scale; thus only in specific astrophysical surroundings are the physics of small and large BHs qualitatively different.

I have chosen to discuss three separate categories of black
holes: those that arise from stellar collapse, those that may
come about from merging and accretion processes and are termed
supermassive black holes (SMBH), and those which may be
relics of the dense highly inhomogeneous medium in the early
universe (primordial black holes---PBH).  

No longer can it be claimed that ``for astrophysical purposes a BH is a Newtonian point mass''; GR has become indispensable to understand fine points of the observations relating BHs.  Yet GR is considered by many opaque and hard to wield.  Thus many model builders use GR formulae and results without really understanding where they come from.  I try to alleviate the situation by providing in these lectures a little primer to GR together with some example calculations which are both easy to do and have important consequences.  I also discuss some concrete issues in BH astrophysics, using whenever feasible GR results here obtained.
 
\subsection{How do black holes form from stars ?}
\label{sec:how}

A good review is that of Brown, et al.~\cite{brown}.  A normal star is stable so long as nuclear burning in it provides thermal pressure to support it against gravity.  But nuclear
burning gradually converts the star core's  hydrogen to helium,
and for massive cores ($M>5M_\odot)$ the helium is
subsequently burned to carbon, and carbon to heavier elements
until we reach the iron group.   The core contracts as each type
of fuel runs out because it temporarily loses pressure until the
heating resulting from the contraction ignites the next type of
fuel.  Since nucleosynthesis of heavy nuclei from lighter ones raises the mean particle mass, it contributes to pressure loss.
 
In the end the endothermic disintegration of iron-group nuclei, the tightest bound of nuclei, precipitates core collapse.  Evolutionary calculations~\cite{brown} show that single stars with initial masses $20-30 M_\odot$ leave cores of more than $1.8
M_\odot$.  These promptly collapse to ``massive BHs''
($M>1.8 M_\odot$) with no optical display, but intense neutrino emission from neutronization.  For initial mass $18-20 M_\odot$ the core implosion (also accompanied by neutronization) engenders a shock wave which can blow part of the star out (type II supernova).  A core of mass $1.5M_\odot<M<1.8M_\odot$, momentarily neutron rich, is left; after deneutronization it collapses into  a ``low mass BH''.  
Stars with initial masses $10-18 M_\odot$ would seem to
make supernovae which leave neutron star remnants.    The fate
of single stars in the initial mass range $\sim 35-80 M_\odot$
is uncertain.

Binaries develop differently.  According to various authors,  a star with initial mass
$20-35 M_\odot$ in a binary can loose its H
envelope by overflowing its Roche lobe to become a ``naked''
helium star, which upon going supernova leaves a low mass BH or a neutron star.   Since a goodly fraction of stars occur in binaries, this is a way to make normal star-BH binaries, provided the BH formation does not disrupt the binary, e.g. by copious gravitational waves emission~\cite{bekenstein}.  The neutron stars in binaries provide a two-step path for black hole binary (BHB) formation following Roche lobe overflow by the the companion star as it evolves to a giant.  Accretion may take the neutron star's mass
over the critical mass, whereupon it will implode to a BH.  Of course, the residual normal star can itself go supernova forming a two-compact object binary. BH-neutron star binaries will be more common than neutron star-neutron star ones, but harder to see.
Stars with $M>80 M_\odot$ in binaries will make high mass BHs (as defined above).  This is the origin of BHBs like Cyg-X1
whose ultimate fate is supernova explosion of the massive
normal star and merger of the compact binary. 

Evidently late stellar evolution yields BHs.  It is estimated that the galaxy contains some $3\times 10^8$ stellar-mass BHs~\cite{mcclintock}.  To discuss BHs we need full GR.

\subsection{The gist of General Relativity}\label{sec:physics}

GR conceives gravity not as a force between masses but as a change in the geometry of {\it spacetime\/} due to neighboring energy.   Consider Euclidean space which may be written in the familiar forms
\begin{eqnarray} 
ds^2&=& dx^2+dy^2+dz^2
\label{euclidean}
\\
ds^2 &=&  dr^2 + r^2 (d\vartheta^2 +
\sin^2\vartheta\, d\varphi^2), 
 \label{phyt}
\end{eqnarray}
or generically as
\begin{equation} ds^2=\sum_{i,j=1}^3 g_{ij}\, dx^i\, dx^j,   
\label{3-D}
\end{equation} 
where the $g_{ij}$ coefficients are called the metric
coefficients.  In principle there are six of them ($g_{ij}$ counts the
same as $g_{ji}$), but in the popular versions of the Euclidean line element, three vanish.
If the six $g_{ij}$ are arbitrary functions, then (\ref{3-D}) does not necessarily describe Euclidean
space, but some curved space in three dimensions, a Riemannian space.   A clear example is the line element $ds^2=d\vartheta^2 +\sin^2\vartheta\, d\varphi^2$ for
two-dimensional space.  It describes the geometry of the surface
of an ordinary ball of unit radius.  This is a curved space (try
flattening a world globe without distorting it) and it is so in
two-dimensions.

In special relativity we deal not just with space, but with flat four-dimensional spacetime, Minkowski spacetime, with metric 
\begin{equation} ds^2=-c^2 dt^2 + dx^2+dy^2+dz^2. 
\label{minkowski}
\end{equation} 
If we replace the space part here by expression (\ref{3-D}), we still deal the Minkowski spacetime. In fact
\begin{equation} ds^2=\sum_{\alpha,\beta=0}^3
g_{\alpha\beta}\, dx^\alpha\, dx^\beta,  
\label{4-D}
\end{equation} 
now with Greek indices like $\alpha$ and $\beta$ running over $0, 1, 2, 3$ ($x^0$ represents the time coordinate, $x^i$ a space one), is still the spacetime of special relativity, provided we came by the $g_{\alpha\beta}$ coefficients by transforming the coordinates
from $t, x, y, z$ in the line element~(\ref{minkowski}).  (Henceforth we use the Einstein convention: any product or single factor where a pair of up and down indices are equal means a sum over this index.  Thus the last formula will be written $ds^2=
g_{\alpha\beta}\, dx^\alpha\, dx^\beta $).  However, for a collection of ten arbitrary functions $g_{\alpha\beta}$ ($g_{\alpha\beta}$ counts the same as $g_{\beta\alpha}$), the spacetime will be curved or (pseudo-)Riemannian spacetime. 

Einstein proposed that each curved spacetime represents a
particular gravitational field.  Thus writing the equations of
nongravitational physics, e.g. Maxwell's, in such a spacetime
rather than in Minkowski spacetime (\ref{4-D}) automatically takes care of the effects of gravity on the physics.   This is now called the Einstein equivalence principle.   Einstein got this revolutionary idea from the universality of free fall: all objects that fall freely in a given gravitational field from identical initial position and velocity move on identical worldlines regardless of differences in structure and composition.  This fact is now tested experimentally to precision of one part in $10^{13}$) (upcoming satellite experiments should improve the number by 3-4 orders).  

Contrast this with motion in an electromagnetic field: a positive particle moves one way, a negative one another way, and a neutral particle with a dipole in yet another.  What happens in gravity seems a big coincidence from this perspective. Einstein elevated the universality of free fall in gravity to the status of a principle: the {\it weak\/} equivalence principle.  He realized that the irrelevancy of structure and composition can be understood if particle worldlines are some universal trajectories in a curved geometry, so that gravity is no longer a force but a reflection of the curvature of spacetime.  Then he generalized the notion to all nongravitational physics---this is the Einstein equivalence principle.  Obviously the two principles are consistent with each other.

What are the universal trajectories ?  In special
relativity a particle not subject to forces moves inertially: if $x^\alpha(\tau)$ is the trajectory $\{x, y, z, t\}$ as functions of the proper time $\tau$, then
\begin{equation} {d^2x^\alpha\over d\tau^2}=0.  
\label{inertial}
\end{equation} 
Now, a theorem in Riemannian geometry tells us that {\it
locally} any metric (\ref{4-D}) with the correct signature can be
rewritten as (\ref{minkowski}) by an appropriate change of
coordinates.  At different points we use different transformations of
coordinates, but always end up with the Lorentz metric in the
new coordinates.  So the equation (\ref{inertial}), when written
in terms of the coordinates for which the metric looks like
(\ref{4-D}), must describe the trajectory in the gravitational
field.  This is the geodesic equation (sum over $\beta$, $\gamma$)
\begin{equation}  {d^2x^\alpha\over d\tau^2}
=-\Gamma^\alpha_{\beta\gamma} {dx^\beta\over d\tau}\,
{dx^\gamma\over d\tau }    
\label{geodesic} \end{equation} with the functions
$\Gamma^\alpha_{\beta\gamma}$ determined uniquely by
($\partial_\alpha\equiv \partial/\partial x^\alpha$)
\begin{equation}
g_{\mu\alpha}\Gamma^\alpha_{\beta\gamma}
={\scriptstyle 1\over \scriptstyle 2}(\partial_\gamma
g_{\mu\beta}+\partial_\beta
g_{\gamma\mu}-\partial_\mu g_{\beta\gamma}).
\label{christofel}
\end{equation} 
Note that the ``gravitational force'' [r.h.s. of (\ref{geodesic})] is velocity dependent.

But is Newtonian gravitation grossly wrong ?  Not at all.  It is
relevant for nonrelativistic motion, namely one with
$|dx^i/d\tau|\ll c$.  This means the terms on the r.h.s. of
(\ref{geodesic}) with a factor
$dx^i/d\tau$ or two must be negligible.  We are left with (no sum; nonrelativistically $\tau\rightarrow t$)
\begin{equation}  {d^2x^i\over d\tau^2} \approx
-\Gamma^i_{tt} {dt\over d\tau}\, {dt\over d\tau } \approx
-\Gamma^i_{tt},      
\label{geoapprox}
\end{equation} 
 where the residual force is velocity independent.  Now
whenever $g_{\alpha\beta}$ is time independent, e.g. field of the sun or a galaxy, and close to the Minkowski metric
 (\ref{minkowski}), we have from (\ref{christofel}) $\Gamma^i_{tt}=-{\scriptstyle 1\over
\scriptstyle 2}\partial_i  g_{tt}$. Thus (\ref{geoapprox}) looks
like the Newtonian equation for motion in a potential
$\Phi_N$ provided
$g_{tt}=-2\Phi_N+$ const.  Comparing with the special
relativistic metric (\ref{minkowski}) we see that nonrelativistic motion
is Newtonian if we can write
\begin{equation} ds^2=-(c^2+2\Phi_N)dt^2 +2\sum_i g_{ti}
dt\, dx^i+\sum_{ij} g_{ij} dx^i\, dx^j.
\label{newtonian}
\end{equation} 
Thus Newtonian dynamics fixes the form of
$g_{tt}$, at least to first order in $\Phi_N$ which is determined by Poisson's equation.  We may rely on the above so long as
$|\Phi_N|\ll c^2$; otherwise the condition
$|dx^i/d\tau|\ll c$ would break down.  Newtonian
dynamics gives no information about  $g_{ti}$ and $ g_{ij}$ save that they cannot be too large.

Let us use (\ref{newtonian}) to compute the
gravitational redshift, the reduction in the
frequency of waves as they climb out of a gravitational
potential well.  Recall that we obtained (\ref{newtonian}) by
assuming $g_{\alpha\beta}$ is time independent.  Imagine an
oscillator (decaying atom, radar device) produces a wave train
of sharp frequency $\nu_1$ at a point ${\bf x}_1$.  This means
that
$N\equiv \nu_1\Delta\tau_1$ is the number of cycles of  the
wave in an interval $\Delta\tau_1$ of the (proper) time ticked
by a clock at rest at 
${\bf x}_1$.  But by (\ref{newtonian}) we have the relation
$\Delta\tau_1=(c^2+2\Phi_N({\bf x}_1))^{1/2}\Delta t$ with
the interval of $t$ time spanned by the train.   Thus the number
of cycles can be written
$N=\nu_1(c^2+2\Phi_N({\bf x}_1))^{1/2}\Delta t$.  Now the
metric is not changing, so $\Delta t$ is also the interval of $t$
time spanned by the wave train anywhere, in particular  at the
destination ${\bf x}_2$.  But wave cycles cannot get lost, so
using {\it his\/} clock, the observer at ${\bf x}_2$ must assign
the wave a frequency $\nu_2$ such that
$\nu_2(c^2+2\Phi_N({\bf x}_2))^{1/2}\Delta t=N$.  We thus
have
\begin{equation} {\nu_1\over \nu_2}={ (c^2+2\Phi_N({\bf
x}_2))^{1/2}
\over (c^2+2\Phi_N({\bf x}_1))^{1/2}}\approx
1+\Delta\Phi_N/c^2 
\label{nuration}
\end{equation} with $\Delta\Phi_N\equiv \Phi_N({\bf
x}_2)-\Phi_N({\bf x}_1)$.   Now the redshift $z$ suffered by
the wave is defined by
$1+z=\lambda_2/\lambda_1=\nu_1/\nu_2$;  hence Einstein's celebrated result,
\begin{equation}  z\approx\Delta\Phi_N/c^2.  \label{1plusz}
\end{equation}

The redshift is positive (negative)  if the waves propagated uphill (downhill) in the
potential, and its magnitude is just the change in
Newtonian potential measured in units of $c^2$.  Formula 
(\ref{1plusz}) applies to the frequency of any repetitive
phenomenon, e.g. the rate of arrival of pulses from a
pulsar.  Applications are many.  The ratio of mass to radius of white
dwarf stars can be deduced from the redshift of the spectral
lines using (\ref{euclidean}).  Of course one has to first correct
for Doppler shift coming from motion along the line of sight.  In
the approximation which we work in, Doppler shift and
gravitational redshift are additive.  Another example more
germane to this lecture: positronium annihilation gives rise to a
gamma ray line at $511$ KeV.  But if the annihilation occurs in
the near environment of a gravitating body, neutron star or black
hole, those photons will be received on Earth with a lower
energy, the defect measuring the depth of the potential well. 

How big can the gravitational redshift get ?
A neutron star of $1.5 M_\odot$ has a radius of $\approx 10$
km giving a formal surface Newtonian potential $\approx
0.22c^2$.  This already calls for nonlinear corrections.   A BH, being
totally collapsed, is more extreme.  We observe from (\ref{1plusz})
that $\nu_2\rightarrow 0$ for any $\nu_1$ when
$\Phi_N({\bf x_1})\rightarrow -c^2/2$.  We are obviously
pushing our formulae too far because they were obtained from
nonrelativistic arguments, and when $\Phi_N$ is of order
unity, motion is relativistic.  Nevertheless the prediction that
the formal Newtonian potential can only reach down to $
-c^2/2$ turns out to be correct, and so is the conclusion that the redshift is then infinite (see below). 

The deflection of light by gravitating bodies is another famous
phenomenon.  Unlike the gravitational redshift, it does depend on $g_{ti}$ and $g_{ij}$.  The reason is that photons move at speed $c$, so the terms in Eq.~(\ref{geodesic}) involving
$\Gamma^i_{tj}$ and $\Gamma^i_{jk}$ are no longer small
compared to that with $\Gamma^i_{tt}$, which was the only
one implicated in nonrelativistic motion.  $\Gamma^i_{tj}$ and
$\Gamma^i_{jk}$ both involve spatial derivatives of
$g_{ij}$, so $g_{ij}$ must be known to calculate light deflection.  Newtonian arguments cannot help us here.

We now turn to Einstein's full gravitational
equation.  There being ten metric components, there are ten
partial differential equations to determine them.  One is a
fanciful elaboration of Poisson's equations with the
relativistic energy density---as opposed to rest mass
density---as source.  Pressure and energy fluxes become the
sources of the others.  If we are mostly interested in the {\it
external\/} gravitational field of a spherically symmetric body,
then the sources can be dropped and the unique exact solution
is Schwarzschild's metric (not Martin Schwarzschild but his
dad Karl Schwarzschild, also the father of photographic
photometry):
\begin{equation} ds^2 = -(c^2+2\Phi_N)dt^2 +
(1+2c^{-2}\Phi_N)^{-1} dr^2 + r^2(d\vartheta^2+\sin^2\vartheta\,
d\varphi^2)  \label{schwarzschild}
\end{equation} As usual, $\Phi_N=-GM/r$ where $M$ is
the mass of the object and $r$ that radial coordinate for which
$4\pi r^2$ is the area of a $r=$ const. surface.  

\subsection{The black hole concept in astrophysics} \label{BH_concept}

Schwarzschild's metric describes the exterior of any spherical mass-energy distribution as well as the simplest BH. 
Obviously the $g_{tt}$ of this metric is the same as for the
approximate metric (\ref{newtonian}).  Hence the discussion
about gravitational redshift of a source at rest goes as before,
and we again have
\begin{equation} z={\nu_1\over \nu_2}-1={
(c^2+2\Phi_N({\bf x}_2))^{1/2} \over (c^2+2\Phi_N({\bf
x}_1))^{1/2}}-1 
\label{z}
\end{equation} But now we can trust the result even when
$\Phi_N$ is not small on scale $c^2$. We confirm that if a source
lies at $r=r_h$, where $\Phi_N=-c^2/2$, then the redshift of radiation it emits is infinite.  Thus the surface at radius
\begin{equation} r_h\equiv {2GM/ c^2}=2.94\ {M/
M_\odot}\ {\rm km},   \label{rhS} \end{equation}  
and more generally any surface where $g_{tt}=0$, is called the surface of infinite redshift.   A fixed light source there is invisible from a distance because its radiation gets shifted to infinite wavelengths.

BHs and other compact objects are frequently surrounded by accretion disks in which gas confined close to a plane moves about the object in nearly circular orbits.   With the Schwarzschild BH in mind we can now understand the phenomenon of asymmetric iron line profiles in the X-ray spectra of BHBs.  Some of these like GRS 1915+105 and V4641 Sgr observed with {\it BeppoSaX\/}~\cite{zand} show the unmistakable Fe K$\alpha$ emission line (from the 2P----1S transition in Fe and its ions) with a broad red tail.  Now, if the disk is not face on, the Doppler shift and relativistic beaming will convert the narrow line into two, the blueshifted component being the stronger.   The second order (relativistic) Doppler effect in conjunction with the gravitational redshift will shift all frequencies down by a fraction ${\cal O}(v^2/c^2)$ (because $v^2/c^2\sim \Phi_N/c^2$).  But light from the inner parts of the disk is shifted more strongly than that from further out, thus also broadening the line.  The three effects together thus broaden asymmetrically about the line's center. As we shall mention, the Schwarzschild accretion disk can extend down to $r=6GMc^{-2}$ so that formula (\ref{z}) attests to a respectable gravitational redshift range quite sufficient to explain some of the mentioned observations.  In other cases, like XTE J1650-500 observed with {\it XMM\ Newton\/}~\cite{miller}, a rotating BH (see next section), whose accretion disk can reach further in, seems to be called for.

We notice that the Schwarzschild $g_{rr}$ metric component blows up at $r=r_h$.  This does not signal physically harsh conditions there, but only that the coordinates $t$ and $r$ do not retain their intuitive meaning at points inside $r=r_h$.  For instance, in that region the path $\vartheta=$ const., $\varphi=$ const. and
$t=$ const. with $r$ varying has $ds^2 = -c^2 d\tau^2<0$, so that it represents a physically possible motion.  But
who ever saw motion with time at a standstill ?  Thus
in the interior region $t$  is {\it not\/} time, that role being
usurped by $r$.   In these lectures we do not care about the
interior, so we can use metric (\ref{schwarzschild}) with the
usual meanings provided we are careful at $r=r_h$.  The surface $r=r_h$ where $g_{rr}$ blows up is the boundary of the BH, and is generally called the {\it horizon\/}. In the Schwarzschild case the roles of infinite redshift surface and horizon are played by the same surface, but this is not a fast rule.
 
The horizon can be crossed by physical entities only inward. 
To see this look at the curve   $\vartheta=$ const.,
$\varphi=$ const. and $r=$ const.  It has $ds^2=-c^2 d\tau^2 = -(c^2+2\Phi_N) dt^2$.  With $r=r_h$, $d\tau =0$, so this is the track of a ``hovering'' photon, and a photon is forever on
the lightcone.  But the lightcone can only be crossed inward.  We conclude that in Schwarzschild spacetime gravity bends
and blends the local light cones at $r=r_h$ into a spherical surface---the  horizon.  Anything inside it is in limbo for it cannot send signals out.  In fact, no characteristic trace of it remains.  A large number of calculations have made it clear that when a nonrotating electrically neutral object totally collapses, the BH, when it settles down, is a Schwarzschild BH with $M$ as its only parameter.  All other information about the object, its chemical composition, its thermal state, etc. is veiled from the exterior.  A BH is the prefect shredder.

Now we apply all this to elucidate the
structure of accretion disks near Schwarzschild BHs.  Ignoring fluid and magnetohydrodynamic effects, we  consider free particles in circular orbits on the equatorial
plane of the Schwarzschild metric.  A key
question is what is the relation between the azimuthal frequency
$\nu_a$ and the radius $r$ of the orbit.  For
example, the frequencies of the quasiperiodic oscillations
(QPOs) seen in some galactic X-ray sources are thought to reflect the $\nu_a$ of close-in orbits.  
 
An easy procedure to $\nu_a(r)$ is as follows.  We set $r$ a constant and  $\vartheta=\pi/2$ (circular orbit in the plane).  The $r$ component of Eq.~(\ref{geodesic}) gives
\begin{equation}
\Gamma^r_{tt}\Big({dt\over
d\tau}\Big)^2+\Gamma^r_{\varphi\varphi}\Big({d\varphi\over
d\tau}\Big)^2+2\Gamma^r_{t\varphi}\Big({dt\over
d\tau}\Big)\Big({d\varphi\over d\tau}\Big) = 0.
\label{free}
\end{equation} The factor $2$ comes about because 
$\Gamma^r_{t\varphi}=\Gamma^r_{\varphi t}$.     Now
since the metric (\ref{schwarzschild}) is $\varphi$ and $t$
independent,  and $g_{\alpha r}\propto\delta_\alpha^r$, it is
evident from Eq.~(\ref{christofel}) that $\Gamma^r_{t\varphi}=0$, while $\Gamma^r_{tt}/\Gamma^r_{\varphi\varphi}=\partial_r
g_{tt}/\partial_r g_{\varphi\varphi}$.  Thus
$\nu_a=(2\pi)^{-1}\Omega$ where
\begin{equation}
\Omega=d\varphi/ dt=(-\Gamma^r_{tt}/
\Gamma^r_{\varphi\varphi})^{1/2}=(-\partial_r
g_{tt}/ \partial_r g_{\varphi\varphi} )^{1/2}.
\label{OmegaS}
\end{equation} 
Since $\sin\vartheta=1$, a simple calculation using Eq.~(\ref{schwarzschild}) gives
\begin{equation}
\nu_a=(2\pi)^{-1}\Big({GM\over r^3} \Big)^{1/2}
=5.79\times 10^{4} 
\Big({M\over M_\odot}\Big)^{1\over 2}
\Big({1\ {\rm km}\over r}\Big)^{3\over 2}\ {\rm Hz}.
\label{fS}
\end{equation}

Coincidentally, this has the the same form as the Newtonian azimuthal frequency. Other (usually smaller) frequencies associated with the orbit are the radial epicyclic frequency $\kappa$ (the frequency of a radial perturbation of Eq.~(\ref{geodesic}) about the circular orbit just discussed), and the vertical frequency $\nu_\perp$ (of perturbations off the exactly circular planar orbit).   The orbit must lie at $r>r_h$; in addition all circular orbits with
$r_h<r<3r_h$ are known  to be unstable: $\kappa\rightarrow 0$ at $r=3r_h$ which is thus known as the innermost stable
circular orbit (ISCO).  Periodic phenomena associated with circular motion must take place outside the ISCO.    Thus Eq.~(\ref{fS}) implies for the frequency of a periodic phenomena,
\begin{equation}
\nu<2.2 (M_\odot/M)\,{\rm kHz}.
\label{largest}
\end{equation}
Interestingly enough, QPOs in galactic X-ray sources have
frequencies from a few Hz reaching up to half a kHz,
corresponding to the largest frequency allowed by (\ref{largest})
for a BH with $1.5$ up to a few solar masses, or to a
solar mass neutron star but with $r$ well outside ISCO (neutron star radii are a few times $r_h$).  These high frequencies are one more proof of the existence of compact objects in nature: even for a white dwarf $\nu_a$ for $r$ outside the star is way too low to fit many of the observed QPOs.

\subsection{Rotating black holes} 
\label{sec:rotating}
The Schwarzschild BH does not rotate.  BH physics tells us that a collapsed neutral rotating star gives a Kerr BH.  Its line element~\cite{MTW} is parametrized by mass $M$ and angular momentum {\it per\ unit\ mass\/} $a$:
\begin{eqnarray} 
ds^2&=&g_{tt}dt^2+2g_{t\varphi}dt
d\varphi+g_{\varphi\varphi}d\varphi^2+
\Sigma\Delta^{-1}dr^2+\Sigma d\vartheta^2
\label{Kerr}
\\ g_{tt}&=& -(c^2-2GMr\Sigma^{-1})
\label{Kerr1}
\\ g_{t\varphi}&=&
-2GMac^{-2}\Sigma^{-1}r\sin^2\vartheta
\\ g_{\varphi\varphi}&=&[(r^2+a^2c^{-2})^2 -a^2
c^{-2}\Delta \sin^2\vartheta ]\Sigma^{-1}\sin^2\vartheta
\\
\Sigma&\equiv& r^2+a^2 c^{-2}\cos^2\vartheta
\\
\Delta&\equiv& r^2-2GMc^{-2}r+a^2c^{-2}.
\label{Kerr2}
\end{eqnarray} 
  Note that the element $g_{t\varphi}$ no longer
vanishes.  The Kerr parameter $a c^{-1}$ has dimensions of length.  The larger the ratio of this scale to $GMc^{-2}$ (the {\it spin\ parameter\/} $a_*\equiv ac/GM$), the more aspherical the metric.   Schwarzschild's BH is the special case of Kerr's for $a=0$.

The infinite redshift surface, this time oblate, is still where $g_{tt}=0$:
\begin{equation} r=r_\infty(\vartheta) \equiv
GMc^{-2}+[(GMc^{-2})^2-a^2c^{-2}\cos^2\vartheta]^{1/2}.
\label{rinf}
\end{equation} Again, radiation reaching a distant observer from an  emitter at rest there has its frequency gravitationally redshifted to zero.  

The horizon, that surface which cannot be crossed outward, is
delineated by the condition $g_{rr}\rightarrow\infty$.  It lies at
$r=r_h$ where
\begin{equation}  r_h \equiv GMc^{-2}+[(GMc^{-2})^2-a^2
c^{-2}]^{1/2}.
\label{rh}
\end{equation} 
Indeed, the track $r=r_h$, $\vartheta=$ const. with $d\varphi/d\tau=a(r_h{}^2+a^2)^{-1}\, dt/d\tau$ has $d\tau=0$ (it represents a photon circling azimuthaly {\it on\/} the horizon, as opposed to hovering at it).  Hence the surface $r=r_h$ is tangent to the local lightcone.   The horizon meets the infinite redshift surface at the poles, but is otherwise inside it. Eq.~(\ref{rh}) should not fool us;  the (Boyer-Lindquist) coordinates used in Eq.~(\ref{Kerr}) make
the horizon {\it look\/} spherical, but it is not because the angular part of the Kerr metric cannot be put in the form $f(r)(d\vartheta^2 +\sin^2\vartheta\, d\varphi^2)$ reflecting the full symmetry of a sphere.  However,  all properties of the metric, including the horizon's shape, are {\it axially\/}
symmetric, and of course, stationary.    The horizon radius $r_h$ is well defined only for $a_*\leq 1$; a BH's angular momentum has a maximum value that rises as $M^2$.

Inside the so called {\it ergosphere}, $r_h<r<r_\infty(\vartheta)$, no particle, whether free-falling or propelled, can stay at fixed $r$, $\varphi$ and $\vartheta$: such track would have $ds^2>0$.  At the very least the particle must circulate constantly in the
angular directions.   This reflects the phenomenon of ``frame
dragging'', common to all axially symmetric metrics with
$g_{t\varphi}\neq 0$.  Contrary to expectations from Mach's principle, the local Lorentz inertial frames are rotating with respect to the inertial frame at infinity (the frame of the stars), and when this phenomenon gets intense enough, it does not let the particles stay in one place.  For large $r$, $g_{t\varphi}\sim r^{-1}$ while both $g_{\vartheta\vartheta}$
and $g_{\varphi\varphi}$ {\it grow\/} as $r^2$, so that the dragging
becomes rapidly imperceptible with growing $r$.
It is strong only in or near the ergosphere.

To get a better feeling for the nature of the frame dragging, let
us find the angular velocity $\Omega\equiv d\varphi/dt$ of a
free particle in a circular orbit $r=$ const. and $\vartheta=\pi/2$ about the BH.  Symmetry tells us such an orbit will not get out of the equatorial plane.  The geodesic equation (\ref{geodesic}) again gives Eq.~(\ref{free}).  In view of definition (\ref{christofel}) we now have
\begin{equation} (\partial_r g_{tt}+2\Omega\partial_r
g_{t\varphi}+ \Omega^2 \partial_r g_{\varphi\varphi})(dt/d\tau)^2=0,
\end{equation} 
which quadratic condition for $\Omega$ has the
solutions
\begin{equation}
\Omega={(GM)^{1/2}\over (GM)^{1/2} ac^{-2}\pm r^{3/2}  }\ .
\label{OmegaK}
\end{equation}

As $a\rightarrow 0$, the $+$ solution here asymptotes to Schwarzschild's (\ref{OmegaS})-(\ref{fS}), as expected.  The
$-$ solution  obviously corresponds to a retrograde circular orbit since it formally gives $\Omega<0$. For $a=0$ both prograde and
retrograde orbits for the same $r$ have the same
$\nu_a=(2\pi)^{-1}|\Omega|$.  But for $a\neq 0$ 
the retrograde orbit has a larger $\nu_a$
than the prograde one with like $r$ !  This effect, totally
foreign to Newtonian theory, is a consequence of the frame dragging phenomenon.  With (\ref{OmegaK}) goes the formula
\begin{equation}
\nu_{a\pm}={3.24\times 10^4(M_\odot/M) \over
(c^2r/GM)^{3/2}\pm a_*}\ {\rm Hz}.
\label{fK}
\end{equation}

\begin{figure}[ht]
\includegraphics[width=2.5in]{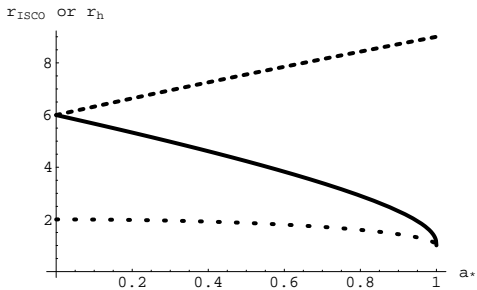}
\vskip.1in
\caption{{\bf The ISCO.\/}  The solid (broken) curve gives the radius of the ISCO of a Kerr BH as function of $a_*\equiv ca/GM$ for prograde (retrograde) orbits, while the dotted line is the radius of the horizon.  All radii are expressed in units of $GMc^{-2}$.}
\label{fig:i}
\end{figure}

Just as in the Schwarzschild case, a Kerr BH has
an ISCO corresponding to the vanishing of the 
epicyclic frequency $\kappa$.   (Both $\kappa$ and $\nu_\perp$ are plotted against $r$ in Ref.~\cite{psaltis}; they are smaller than $\nu_a$.)  The formulae for the innermost stable orbit radius $r_{\rm ISCO}$ for both prograde and retrograde orbits are given by Bardeen et al.~\cite{bardeen}, and are here plotted in Fig.~\ref{fig:i} together with the horizon radius $r_h$, all as a function of $a_*$.  The large value of $r_{\rm ISCO}$ for
retrograde orbits is what prevents the
formal pole in Eq.~(\ref{fK}) from showing up as
$a_*\rightarrow 1$ when $r_h\rightarrow GMc^{-2}$.

\begin{figure}[ht]
\includegraphics[width=2.5in]{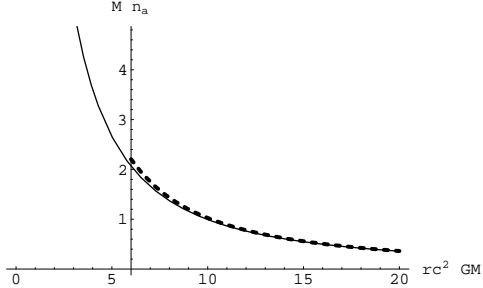}
\vskip.1in
\caption{{\bf The azimuthal frequency.\/}  The azimuthal frequency $\nu_a$ in kHz multiplied by $M/M_\odot$ for prograde circular orbits outside ISCO in the equatorial plane of a Schwarzschild BH (dashed) and extreme ($a_*=1$) Kerr BH (solid curve) as function of $r c^{-2}/GM$. }
\label{fig:fa}
\end{figure}

Case $+$ of formula (\ref{fK}) is plotted in Fig.~\ref{fig:fa} for the stable circular orbits around a Schwarzschild and extreme  Kerr ($a_*=1$) BH.  It may be seen that in range of $r$ where they overlap, the curves are quite similar.  However, for a rapidly spinning BH the curve extends much further in.

\subsection{QPOs in X-ray binary black holes}

In a detailed review of BHB, McClintock and Remillard~\cite{mcclintock} list some 40 well observed X-ray binary sources in the galaxy with candidate BHs, about half of the number with high confidence.  These BHBs, often called microquasars, exhibit five distinct spectral/temporal states, each source flipping among several of these:  (1) the thermal--dominant (TD) or high/soft (HS) state, a high-intensity state dominated by thermal emission from the accretion disk; (2) the low/hard (LH) state, a low-intensity state dominated by power-law emission spectrum with a photon index $\approx 1.7$, rapid variability and frequently accompanied by a radio jet; (3) the quiescent state, of very faint luminosity $3\times 10^{30}-3\times 10^{33}$ erg s$^{-1}$, and also dominated by power--law emission (index 1.5--2.1) whose very faintness is regarded as good evidence that the accreting surface is a horizon and no solid boundary; (4) the very high (VH) or steep power law (SPL) state with index 2.5 or so, and (5) the intermediate state.  The best candidates for BHB have been seen in all of the HS, LH and VH states.  The transition between two different states is thought to be related to changes in the rate of accretion to the disk.

A respectable minority of BHB candidates exhibit bumps in the power spectrum of the X-ray count rate; these are called quasiperiodic oscillations (QPOs; as opposed to periodic which would imply a sharp frequency). Here we are not interested in the low frequency (0.1--few Hz) QPOs seen for days or weeks during the VH state with amplitudes of order 10\% of the total X-ray counts, and whose median frequencies often change with the X-ray flux.  Rather we shall dwell on the high frequency (40--450 Hz) HFQPOs seen in seven of the BHB candidates during the VH state.  These QPOs usurp 1--3\% of the X-ray flux.  Three of them exhibit single frequency QPOs: 4U163047 (184 Hz), XTE J1859+226 (190 Hz), XTE J1650-500 (250 Hz).  Three others have each a pair of HFQPOs: H 1743-322 (240 Hz + 160 Hz), GRO J1655-40 (450 Hz + 300 Hz) and XTE J1550-564 (276 Hz + 174 Hz).  In addition the source GRS 1915+105 possesses {\it two\/} pairs of HFQPOs (168 Hz + 113 Hz and 67 Hz + 41 Hz) which actually show up during the HS state.  These QPO frequencies do not vary with the X-ray flux; they are a fingerprint of the system.

The first four  pairs of frequencies stand in the ratio of 3:2; in H 1743-322 and GRO J1655-40 the accuracy of this ratio is exquisite, in GRS 1915+105 it is better than 1\%, and in XTE J1550-564 about 5\%.   Even more striking, the lower QPO-pair frequency has been found to follow the empirical law $ M\nu_0\approx 931 M_\odot\,$Hz ($\nu_0$ is half that frequency and is identified as the ``fundamental'' frequency) with respect to reliable determinations of the BH masses~\cite{mcclintock}.  What is this telling us ?
   
A suggestive analog is resonances in our planetary system, e.g. the fact that the orbital period of a Saturnian satellite and that of a subset of the particles in the planet's rings stand in the ratio of small integers makes gaps in the ring.   Abramowicz and colleagues~\cite{abramowicz1} suggest that in the twin QPOs one is seeing resonances of the frequencies $\kappa$ and $\nu_\perp$ for orbits at a fixed value of $rc^2/GM$ for various systems.  Recall that the Kerr metric has only one scale, $GM/c^2$, and one dimensionless parameter, $a_*$, and that $\nu_a$'s dependence on $a_*$ is weak (Fig.~\ref{fig:fa}).  Orbits at fixed $rc^2/GM$ around various BHs are thus similar in a geometric sense, and phenomena involving just them could be expected to be alike despite the range of $M$ involved.  

Now whereas in Newtonian gravity $\nu_a$, $\kappa$ and $\nu_\perp$ are identical, they differ for orbits around (Kerr) BHs.  In fact  almost by definition $\kappa$ becomes relatively small as the orbit approaches the ISCO.  So, for example, there are stable orbits at a fixed $rc^2/GM$ around Kerr BHs for which $\nu_\perp/\kappa$ is exactly $3/2$; the resonance could amplify radial and off-plane perturbations, and the consequent ``rattling'' might well get imprinted on the X-ray power spectrum.  Although it is not clear what the physics of the modulation is, the repetition of the $3/2$ strongly suggests that the same resonance shows up in all four BHBs.  Now since by Fig.~\ref{fig:fa} $M \nu_a $ is fixed for given $rc^2/GM$, and $\kappa$ and $\nu_\perp$ are then definite fractions of $\nu_a$, we see that the McClintock-Remillard $\nu_0\propto M^{-1}$ law is clean evidence that we are dealing with QPOs of BHs.

If these ideas are correct, then the azimuthal frequency associated with each QPO pair must obey $M\nu_a > 3\times 0.931 M_\odot\,$kHz (recall that $\kappa$ and $\nu_\perp$ lie below the corresponding $\nu_a$).  Then Fig.~\ref{fig:fa} discloses that the QPO source must lie at $r<5 GMc^{-2}$: QPO pairs are associated with compact objects, and they are inner disk phenomena.  In fact,  $5 GMc^{-2}$ falls below $r_{ISCO}$ for Schwarzschild, so that some QPO pairs are Kerr BH phenomenon.    Abramowicz et al.~\cite{abramowicz1} have estimated $a_*\approx 0.9$ for GRO J1655-40: rapidly rotating BHs exist in nature (more in Sec.~\ref{sec:spin}). 

\subsection{Supermassive black holes in galactic nuclei} \label{nuclei}

Soon after the discovery of QSOs, 
Salpeter~\cite{salpeter} and  Zel'dovich~\cite{zeldovich} suggested accretion onto a supermassive black hole (SMBH) as the QSO energy source.   Lynden-Bell~\cite{lynden-bell} solidified this understanding by stressing that the large energy in a QSO radio halo, $\sim 10^{54}\,$ J, were it produced by nuclear reactions  whose maximal efficiency is .007, would require the ``burning'' of $10^9\, M_\odot$.   This massive a refuse, concentrated within $10^{13}\,$ m (a dimension required by the $\sim $day timescale of QSO X-ray flux variability) speaks for $10^{55}\,$ J of gravitational binding energy.  So whether nuclear power has any role in fueling a QSO or not, gravitational accretion power {\it must\/} be involved.  

The involvement of deep gravitational wells in QSO's was further exposed by the observations of seemingly ``superluminal'' jets emerging from a number of QSOs, jets which are straight for up to a Mpc.    The ``superluminal'expansion'' implies speeds that approach $c$, so the relativistic parameter $GM/Rc^{2}$ in the accelerating region must approach unity~\cite{rees}.  Rotation of a massive ``gyroscope'' is the logical way to stabilize the emission direction so that the jets are straight, and, of course, BHs can be massive rotators~\cite{rees2}.  A good review of SMBHs in AGNs is given by Laor~\cite{laor}.  By the 1980's it was accepted that QSO's are ephemeral displays in nuclei of galaxies, so that if they implicate SMBHs, then SMBHs may be left in all galaxies which once harbored QSOs. 

More recently direct evidence for SMBHs in not obviously active galaxies has emerged.  The evidence for some 40 of them is discussed in an extensive review by Kormendy and Gebhardt~\cite{kormendy}.  The strongest case is linked with the radio source Sgr A* at the center of the Galaxy.  A cluster of stars is observed within 0.02 pc of it, and motions of some of them with velocities up to 1350 km s$^{-1}$ are seen to change in time; the stars orbit the radio source in tens of years !  Well determined  acceleration vectors for several of the stars  point at Sgr A* and the data allow a determination of the central mass at $(2.6\pm 0.2)\times 10^{6}\, M_\odot$~\cite{genzel}.  This massive object cannot itself be a cluster of smaller objects:  $10^6\, M_\odot$ of ordinary stars would be visible; were it brown dwarfs, they would be so densely packed that they would collide, merge and become luminous beyond observable limits. Finally a $10^6\, M_\odot$ cluster of collapsed stars would be rapidly whittled away by evaporation~\cite{maoz}.

Another strong case is in the nucleus of the spiral NGC 4258 where a number of H$_2$O masers are observed.  The radial (Doppler) velocities can be fit by rotation of a circumnuclear disk of pc scale with a Keplerian rotation velocity profile reaching up to 1080 km s$^{-1}$.  This means there is a ``point'' mass at the center to the tune of  $4\times 10^7\,M_\odot$.  Once again we would be hard put to squeeze this massive a cluster of brown dwarfs or collapsed objects inside the disk's inner radius at 0.13 pc~\cite{maoz}.  It must thus be a BH.  The Seyfert nuclei of NGC 1068 and the Scd NGC 4945 also have (more modest) SMBHs discovered in them by the maser method.

\subsection{Supermassive black holes as energy sources}\label{energy_sources}

Salpeter, Zel'dovich and Lynden-Bell all stressed the great efficiency with which an accreting Schwarzschild SMBH produces radiation.  Later Bardeen~\cite{bardeen2} emphasized that BHs in nature are likely to be rotating, so that one should consider the energy efficiency of an accreting Kerr BH.  Let us use the BH physics we have learned to compute the maximum such efficiency.  The physical picture is of an accretion disk in which matter slowly spirals inward.  As we saw, the spiraling stops at $r_{ISCO}$ where an instability develops, circular motion is no longer possible, so that the disk ends there, if not sooner.  

First we establish that provided gas dynamics and MHD effects are negligible, the correct expression for relativistic total energy per unit mass of a gas parcel is $E=-g_{tt}dt/d\tau-g_{t\varphi}d\varphi/d\tau$.  In flat spacetime $g_{tt}=-c^2$ while $g_{t\varphi}=0$, so $E=c^2 dt/d\tau$.  But from special relativity $dt/d\tau$ is the Lorentz $\gamma$, so $E=\gamma c^2$, indeed the relativistic energy of a unit mass free particle.  Returning to curved spacetime we now show that $E$ is conserved in free motion, so that it correctly includes the gravitational energy.  

In fact, whenever a metric is independent of some coordinate $x^\mu$ (time or a spatial coordinate), the quantity $K=g_{\mu\alpha}dx^\alpha/d\tau$ is a constant of the motion for a free particle (no pressure, dissipation, etc.). For
\begin{equation}
{dK\over d\tau}=\partial_\beta\, g_{\mu\alpha}{dx^\beta\over d\tau}{dx^\alpha\over d\tau}-g_{\mu\alpha}\Gamma^\alpha_{\beta\gamma}{dx^\beta\over d\tau}{dx^\gamma\over d\tau},
\label{rate}
\end{equation}
where we have used the geodesic equation (\ref{geodesic}) to get the second term.  But Eq.~(\ref{christofel}) and the assumed symmetry tell us that
\begin{equation}
g_{\mu\alpha}\Gamma^\alpha_{\beta\gamma}
={\scriptstyle 1\over \scriptstyle 2}(\partial_\beta\,
g_{\mu\gamma}+\partial_\gamma\, g_{\beta\delta}).
\end{equation}
And because of the symmetry between dummy indices $\beta$ and $\gamma$, the $\Gamma$ term in Eq.~(\ref{rate}) exactly cancels the first term.  We did not need to know $g_{\alpha\beta}$ explicitly to show this, only that it does not depend on $x^\mu$.  Since Kerr's metric is $t$ independent, and $g_{tr}=g_{t\vartheta}=0$, $K=-g_{tt}dt/d\tau-g_{t\varphi}d\varphi/d\tau$ is conserved in free motion. By analytical mechanics it must be proportional to the energy, and indeed it is just $E$, the proposed expression for energy.

In reality as a gas parcel orbits, dissipation and radiative losses cause its $E$ to decrease, so that it spirals inward.  Thus the $E(r)$ for circular motion in Kerr as given above means the specific energy {\it remaining\/} by the time the parcel reaches $r$.
To reduce $E(r)$ to a practical form recall that $ds^2=-c^2d\tau^2$, so that for circular motion 
\begin{equation}
c^2 d\tau^2=-g_{tt}dt^2-2g_{t\varphi}dt d\varphi-g_{\varphi\varphi}d\varphi^2=-g_{tt}dt^2\left(1+{2g_{t\varphi}\Omega\over g_{tt}}+{g_{\varphi\varphi}\Omega^2\over g_{tt}}\right).
\label{proper}
\end{equation}
Substituting $dt/d\tau$ from here, $d\phi/dt=\Omega$ as given for prograde orbits by Eq.~(\ref{OmegaK}), and the metric elements (\ref{Kerr1}-\ref{Kerr2}), we get after some labor
\begin{equation}
E=c^2{1-2GM/c^2r+a_*(GM/c^2r)^{3/2}\over[1-3GM/c^2 r+2a_*(GM/c^2 r)^{3/2}]^{1/2}}\ .
\label{spec_energy}
\end{equation}

\begin{figure}[ht]
\includegraphics[width=2.5in]{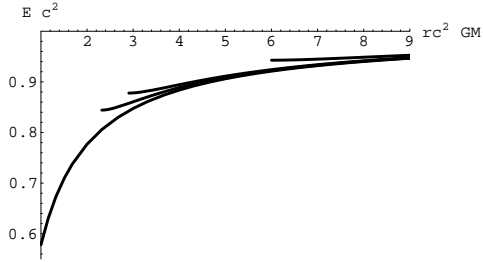}
\vskip.1in
\caption{{\bf Specific energy.\/}  $Ec^{-2}$ per unit mass of a parcel in circular orbit of radius $r$ around a Kerr BH with (in ascending order) $a_*=1,0.9,0.8,0$.  Each curve is shown down to the corresponding ISCO.  Radii are expressed in units of $GMc^{-2}$.}
\label{fig:E}
\end{figure}
Fig.~\ref{fig:E} plots $Ec^{-2}$ for a range of radii down to $r_{ISCO}$ for BHs with several spin parameters.  $E$ starts very near $c^2$ in the outer disk edge because $|\Phi_N|=GM/r\ll c^2$ there: the parcel's energy is then its full rest energy. In line with our previous remark, the distribution of $E$ with radius is a good predictor of how much energy is deposited at any particular ring of the disk; this in turn sets a bound on the emissivity. The steep slope of the curve for $a_*$ shows that much radiation will come from near the ISCO (requiring corrections for losses into the hole).

\begin{figure}[ht]
\includegraphics[width=2.5in]{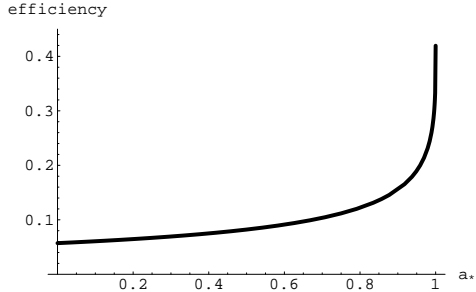}
\vskip.1in
\caption{{\bf Efficiency.\/}  The peak efficiency for conversion of mass to radiation by a disk around a Kerr BH with spin parameter $a_*$.}  
\label{fig:effic}
\end{figure}

At ISCO $E$ reaches its lowest value, for thereafter it cannot spiral in, but must take a plunge into the hole so rapidly that gas dynamical processes make little difference, and $E$ is very nearly conserved thereafter.  Thus $c^{-2}E_{ISCO}$ signifies the fraction of the original parcel rest mass finally accreted by the hole.  Then $1-c^{-2}E_{ISCO}$ is the peak efficiency (``peak''  since some of the disk's radiation may get emitted into the BH).  This is plotted in Fig.~\ref{fig:effic} by taking Fig.~\ref {fig:i} into account in Eq.~\ref{spec_energy}.
For a Schwarzschild BH ($a_*=0$) the efficiency is 0.0572, while for an extreme Kerr BH ($a_*=1$) it reaches 0.42.  Thus a BH accretion disk can convert mass into energy  up to two orders of magnitude more efficiently than nuclear burning: Einstein's $E=mc^2$ at its best !

\subsection{The origin of black hole spin}\label{sec:spin}

But why to expect a central SMBH to be surrounded by an accretion disk ? Gas collecting in the hole's deep gravitational well will come from some distance out.  Whatever angular momentum it has will prevent it from immediately plummeting into the BH, and it will rather swirl around.  Any rotation of the galaxy (even ellipticals and bulges rotate somewhat) will impose coherent swirling of different gas streams, and collisions among them will align the divers angular momentum vectors leading to formation of a thin disk.  

There is direct evidence for accretion disks in AGNs from observations by the X-ray satellite ASCA of the Fe K$\alpha$ line coming from many such sources (and lately also from {\it Chandra\/} and XMM {\it  Newton}~\cite{turner}).  The line profile shows the asymmetry discussed in Sec.~\ref{BH_concept}.  Models which assume a profile for the disk emissivity distribution consistent with the specific energy profile of Fig.~\ref{fig:E} are successful in reproducing the observed shape of the line from the Seyfert nuclei of NGC 3516 and NGC 3227~\cite{nandra}. In the latter case there is some indication that a rapidly rotating BH is involved because the strong asymmetry seen requires very relativistic conditions found only if the disk reaches close to the horizon. 

On this note it is easy to extend the calculations of the preceding section to show why, as Bardeen expected, a BH in an AGN spins rapidly. The physical reason is that the disk feeds the hole with a high angular momentum to mass ratio.  To make matters simple I assume the BH is initially Schwarzschild with mass $M_0$.  We need an expression for specific angular momentum about the axis of the disk.  For nonrelativistic motion in Euclidean space this would be $(r\sin\vartheta)^2\, d\varphi/dt=g_{\varphi\varphi}\Omega$ (see Eq.~(\ref{euclidean})).  For relativistic free motion in Kerr spacetime we should replace this by something similar which is conserved due to the rotational symmetry about the axis (as should angular momentum).  The rule stated just ahead of Eq.~(\ref{rate}) suggests the conserved quantity 
\begin{equation}
\ell=g_{\varphi t}dt/d\tau+g_{\varphi\varphi}d\varphi/d\tau.
\label{ell}
\end{equation}
In the limit of flat spacetime ($g_{\varphi t}\rightarrow 0$) this indeed reduces to the nonrelativistic expression when $\tau\rightarrow t$, so $\ell$ is the specific angular momentum.  

When a unit mass in the disk reaches ISCO, it plummets into the BH without further changes to its energy $E$ and angular momentum $\ell$.  Thus its accretion causes changes $\delta M=E$ and $\delta(aM)=\ell$.  Meanwhile,  a simple calculation gives for an increment of $a_*=ca/GM$,
\begin{equation}
\delta a_* =[c(GM)^{-1}d(aM)/dM-2a_*]M^{-1}\delta M.
\end{equation}
Putting $d(aM)/dM=\ell/E$ and substituting the explicit formulae for $E$ (from Sec.~\ref{energy_sources}), $\ell$ from Eq.~(\ref{ell}) and $\Omega$ from Eq.~(\ref{OmegaK}), and evaluating  at $r=r_{ISCO}$, we obtain the differential equation for the growth of $a_*$ with respect to $M$: 
\begin{equation}
da_*=\left[{\varrho(a_*)^2-2a_*\varrho(a_*)^{1/2}+a_*^2\over \varrho(a_*)^{1/2}[\varrho(a_*)-2]+a_*}-2a_*  \right] d\ln M.
\end{equation}
Here $\varrho(a_*)\equiv c^2(GM)^{-1}\, r_{ISCO}$ is that complicated function of $a_*$ used to make Fig.~\ref{fig:i}.
Integrating this equation numerically from $a_*=0$  at $M=M_0$  shows that the spin parameter reaches $a_*=0.3058$ ($a_*=0.7689$) when $M$ has grown by 10\% (40\%).  By the time the BH has doubled its initial mass, $a_*=0.9875$.  Since $a_*\leq 1$, we see that if a SMBHs has changed its mass substantially by accretion from its disk, it must be rotating rapidly.  Clearly the idealization of an initially Schwarzschild BH is not crucial.  An originally mildly rotating BH would be transformed into an extreme Kerr BH by the time it
gained significant mass. 

\subsection{How are SMBH masses measured ?} 

Most SMBHs have been identified in early type galaxies~\cite{kormendy} (exceptions include the mentioned maser galaxies, the Galaxy and the two nearby Sbs M31 and M81).  No SMBHs have been found in pure disk pr Ir galaxies.  Detecting a BH means finding kinematic evidence (velocities of emission lines from a circumnuclear ionized gas disk, or velocities of absorption lines from bulge stars) for a large mass which is unresolved (more on this below).  Only 10-15\% of early type galaxies have well formed central gas disks, but more than ten have yielded evidence for SMBHs, including the large E0 galaxy M87 which sports a $3\times 10^9\, M_\odot$ hole as well as the famous jet which is perpendicular to the gas disk.  Gas disk derived masses are regarded as accurate to $\sim 30\%$.  Mass determination via stellar velocity dispersion is hampered by the long integration times required (ellipticals and bulges have low surface brightness), as well as by ambiguities of interpretation, e.g. is the velocity distribution anisotropic  ?  Yet more than 25 SMBH masses have been measured by this method.  The range of SMBH masses is $10^6-10^9\, M_\odot$.

The detailed tables~\cite{kormendy} are less important than two correlations that emerge from them.  First, the BH mass $M$ is almost proportional to the bulge blue luminosity $L_b$, or more precisely~\cite{kormendy2,kormendy}
\begin{equation}
M \approx 7.8\times 10^7 (L_b/10^{10}\, L_\odot)^{1.08}.
\label{MvsL}
\end{equation}
And the BH mass is nearly proportional to the fourth power of the bulge velocity dispersion $\sigma_e$ within de Vaucoleurs' effective radius~\cite{ferrarese,kormendy}:
\begin{equation}
M \approx 1.3\times 10^8 (\sigma_e/200\, {\rm km\ s}^{-1})^{3.65}.
\label{Mvs_sigma}
\end{equation}
This last has a scatter not much bigger than the mass measurement errors can account for (see graph in ref.~\cite{kormendy}), so it seems to be the more fundamental.   The SMBH accounts for about 0.0013 of the galaxy's or bulge's mass, but deviations of an order of magnitude are known.  

As a rule SMBHs show up in every galaxy (or bulge) observed with enough resolution to find a BH.  In what sense ?  Suppose that from the observed $L_b$ and $\sigma_e$ an estimate is made of a central SMBH's $M$.  Then the radius of its sphere of influence (region where its mass disturbs stellar orbits drastically) $r_*=GM/\sigma_e^2$ is determined.  Wherever this sphere can be resolved, evidence for a SMBH has always turned up.  Hence the widely accepted belief that most early type galaxies have a central SMBH.  SMBHs in disk galaxies with small bulges---if they exist---fail to comply with the $M-\sigma_e$ relation.  For example, for the Scd M33 a bound of $10^3\, M_\odot$ has been put on any central BH~\cite{kormendy}, yet the $M-\sigma_e$ relation would predict one four orders heavier.  SMBHs go naturally with early type galaxies and big bulges, not with disks.

AGNs are generally too far for the BH sphere of influence to be resolvable.  For them the method of reverberation mapping has been useful.  In a AGN there are rapid optical and UV fluctuations.  Reverberation mapping interprets the time delays between variations in the AGN continuum (thought to come from the accretion disk) and its broad emission lines (coming from enveloping gas clouds where the continuum is reprocessed into lines) as the light travel times between BH and the clouds, and hence equal to the linear size of the region $R$ divided by $c$.   The width of the emission lines gives a velocity dispersion $\sigma$.  Then, assuming that the BH is the main mass in the region one has the virial estimate $M\approx \sigma^2\, R/G$.   This method provides no direct evidence for a point mass.  And it is somewhat model dependent.  Yet, after some initial confusion, it has been agreed that central SMBH reverberation masses lie on the same $M-\sigma_e$ relation as SMBH masses in normal galaxies as determined by the gas disk and stellar dynamical methods.

\subsection{Where did the SMBHs come from ?}

Like questions of heritage in everyday life, the origin of the SMBHs abounds in controversy.  Did SMBHs precede their galaxies and act as nuclei for their formation \cite{ryan,rees2} ?  Or did they grow from humble seeds to their present proportions by accretion from an already developed bulge or elliptical ?  Or did each develop together with its host galaxy ?  The dearth of SMBHs in pure disks suggests that a BH is not required for galaxy condensation, thus militating against the first hypothesis.  On the other hand, their presence in late-type systems is reasonable if SMBHs form by accretion over time; in disks rotational support would impede their growth, and so their absence is also reasonable. 

Early type disk galaxy {\it bulges\/}, like ellipticals, obey the Faber-Jackson relation~\cite{FJ} $M\sim \sigma^4$ between mass and velocity dispersion.  Relation (\ref{Mvs_sigma}) has very much this form; hence as hinted at earlier, a SMBHs tends to contain a fixed mass fraction of its elliptical or bulge host.  This again seems to support the joint growth hypothesis; for how could the full blown SMBH manage to control its host's final mass despite the small dimension of the former's sphere of influence.  SMBHs have been found in AGNs out to $z=6$, some with masses $\sim 10^9\, M_\odot$~\cite{fan}.  Evidently these too militate against the idea of late SMBH formation in fully blown galaxies.  As mentioned, these early SMBHs obey the same rule (\ref{Mvs_sigma}) as ``normal'' SMBHs at the present epoch, suggesting that they too developed together with their hosts.
 
Accordingly, Kormendy and Gebhardt~\cite{kormendy} reach the verdict that AGN activity, bulge formation and SMBH gestation are facets of one and the same primeval process.  This granted, there are remain unsolved problems regarding the path by which BHs grow to SMBH size, as well as regarding the nature of the seed BHs at the outset of the evolution.   Two categories of growth mechanism stand out in the literature: accretion by BHs~\cite{rees1} and mergers of BHs~\cite{barkana}.  Seed candidates include stellar-mass BHs, intermediate-mass ($10^2-10^5 M_\odot$) black holes (IMBHs) and PBHs.  Any seed-growth method combination has to be judged by its success in producing the $10^9\, M_\odot$ SMBHs in AGNs at $z>6$.

A stellar-mass BH accretes slowly.  Early type galaxies are gas poor, but have stars.  To accrete a star the relatively tiny BH must first capture it, and then tidally disrupt it in order to ingest it, all the time respecting the Eddington limit.  It is known that this process acting in early galaxies would have trouble producing $10^9\, M_\odot$ SMBH by $z=6$~\cite{yoo}.  But stellar-mass black holes could play out their role in a massive star cluster.  Such assembly would tend to drop to the bottom of the galactic potential well by dynamical friction on the background, lighter, stars.  Simultaneously distant encounters among its constituent stars would cause it to develop a tightly bound core and a loosely bound envelope; the last would slowly be lost to evaporation.  

In the dense core stellar collisions would be common and would lead to stellar amalgamation, and the consequent formation of massive short-lived stars which would evolve rapidly, some leaving BHs~\cite{rees3,quinlan}.  Of course collisions between these last and the remaining stars should make the holes larger.  In the meantime some of the lighter objects of either kind would be expelled from the cluster by Newtonian slingshot.  Consolidation of the cluster into an IMBH by runaway evolution would follow~\cite{quinlan}.  Other studies~\cite{baumgardt} are more pessimistic regarding the possibility of producing IMBHs by cluster evolution in one Hubble time.

Elliptical galaxies (and bulges) are thought to have grown by mergers of smaller building blocks at $z\sim 3$.  Being very massive on stellar scale, the clusters (or their IMBH remnants) would be caused to congregate at the merged galaxy's center by dynamical friction.  Their mergers would thus be sped up ultimately forming baby SMBH ~\cite{rees3}. However, this mechanism faces a challenge.  Violent formation of a black hole can lead to its strong recoil by the emission of gravitational waves; this process can expel fresh stellar-mass BH's from the Galaxy~\cite{bekenstein}.  It was recently realized that it can also expel newly merged SMBHs from their host bulges or galaxy center's~\cite{favata}, thus impeding their growth.

\subsection{Primordial black holes}\label{sec:PBH} 
 
These problems have riveted attention on the notion that BHs originating in the early universe are the seeds from which SMBHs grow~\cite{bean}.  PBHs, thus far conjectural, offer so much that is novel that they deserve a look in their own right.  Because there is as yet no observational confirmation of PBHs, I treat them in brief. Carr~\cite{carr} and Kiefer~\cite{kiefer} are recent reviews on PBH astrophysics (the second on BH physics as well).

According to Eq.~(\ref{density}) the lighter a BH, the denser:  a $10^9\, M_\odot$ SMBH has a density like snow's; a stellar mass one is above nuclear density.  And a BH the mass of a mountain ($5\times 10^{12}\, {\rm Kg}$) has a whopping density $2.5\times 10^{55}\,{\rm Kg}\, {\rm m}^{-3}$.  Intuitively, to form a BH of such high density should require matter or radiation about as dense.  Only in the early universe do we meet densities orders above nuclear density.  Hence when tiny BHs are the issue, attention turns to cosmology, and they are termed PBHs.

High density is not enough; a highly dense homogeneous distribution of matter$/$radiation is perfectly consistent with cosmology.  To make BHs one needs inhomogeneities.  And obviously collapse cannot take place until the inhomogeneity's radius $R$ has fully entered the (particle) horizon (radius $ct$ with $t$ the cosmological time).  Now during inflation the exponential expansion stretches every $R$ outside the horizon (which grows more placidly); hence PBHs cannot form then.  In addition, any PBHs from former eras are rapidly diluted.  The post-inflation radiation era is the first relevant one in connection with PBHs formation.

As an example take flat space ($k=0$) radiation dominated cosmology; we know that the expansion factor $a\propto \surd t$.  Thus the Hubble parameter is $H={\scriptstyle 1\over\scriptstyle 2}t^{-1}$, and the condition is $R< {\scriptstyle 1\over\scriptstyle 2}cH^{-1}$. But for gravitation to overpower pressure, $R$ must exceed Jeans' length $c_s(4\pi G\rho)^{-1/2}$, where $c_s$ is the sound speed and $\rho$ the mass density.  Due to radiation dominance $c_s=c/\surd 3$, and since Friedmann's equation gives $H^2=(8\pi/3)G\rho$, we have $R>{\scriptstyle 1\over\scriptstyle 3}\surd 2\, c\,H^{-1}=0.47 c\,H^{-1}$.  Comparing we see that the collapse is possible only briefly after horizon crossing; thereafter $R$ becomes very small compared to $cH^{-1}$ and the second condition fails.

From Eq.~(\ref{rhS}) for $R$ and $H\approx {\scriptstyle 1\over\scriptstyle 2}t^{-1}$ we have for the time of formation of a PBH of mass $M$ ($M_*\equiv 10^{12}\,{\rm Kg}$):
\begin{equation}
t\approx 4.9\times 10^{-24}(M/M_*)\,{\rm s}
\label{t}
\end{equation}
As mentioned this is relevant for post-inflation, that is $t>10^{-35}\,{\rm s}$ or $M>1\,{\rm Kg}$.   Since the smaller the PBH, the earlier it forms, the smallest PBHs detectable would probe the earliest cosmological times. Once the PBH is formed, $M$ is not increased sizably by accretion~\cite{carr}.  On the other hand, $M$ can decrease by Hawking radiation.

According to BH thermodynamics~\cite{bekenstein2,hawking}, a Schwarzschild BH is endowed with a quantum temperature
\begin{equation}
T_{BH}={\hbar c^3}(8\pi GkM)^{-1}=1.23\times 10^{11} (M_*/M)\ {}^{0}{\rm K}
\label{T}
\end{equation}
where $k$ stands for Boltzmann's constant.  This is manifested by the BH emitting into space any kind of particles in nature, each with thermal spectrum and statistics corresponding to the Hawking temperature $T_{BH}$. We see that for $M\gg M_*$, only photons and neutrinos will be emitted.  Anyway, this radiation is paid for by a decreasing $Mc^2$. Estimating the energy loss with the Stefan-Boltzmann law $P =(4\pi r_h{}^{2}) \sigma T^{4}$ using $T$ from Eq.~(\ref{T}), we get for each  massless (or very light) species
 \begin{equation}  dM/dt\approx -4.0\times
10^{-9}(M_*/M)^{2}\, {\rm Kg\, s}^{-1}.
 \label{dm/dt}
\end{equation}  
In reality, the BH radiation crossection is larger than $4\pi r_h^2$ by a factor of $\sim 5$.  And since there exist very light neutrinos and antineutrinos, the total $dM/dt$ is an order of magnitude above Eq.~(\ref{dm/dt}).  Integrating the fixed-up equation from $M=M_0$ to $M\approx 0$ gives the lifetime $\tau< 27\times 10^{10}(M/M_*)^{3}\, {\rm y}$, the inequality appearing because as $M$ approaches $M_*$, the PBH begins emitting massive particles and  $M$ falls more quickly than just estimated.  Evidently PBHs with $M\ll M_*$ have evaporated by now.  Nevertheless, as we shall see, PBH searches say  a lot about the inflationary era (at whose end PBHs of only $M\sim 1\,{\rm Kg}$ would form). 

PBH formation is bound up with the primordial density fluctuations spectrum.  Inflation produces a scale invariant spectrum: the root mean square fluctuation $\epsilon$ for an inhomogeneity just inside the  horizon is $M$ independent.  When the propitious time (\ref{t}) comes, a PBH can form only if the density contrast $\delta\equiv \delta\rho/\rho$ at horizon scale is large enough; the rule is that for equation of state $p=w\rho$, $\delta$ must exceed $w$. Evidently, the fraction $\beta$ of horizon-size volumes which actually collapse (the probability that the inhomogeneity makes a PBH), is $M$ independent. 

To calculate the present space density of PBH formed during the radiation, we look at a fixed comoving volume of space; it expands as $a^3\propto t^{3/2}$.  By contrast the volume contained in a horizon is $\propto t^3$: the large volume thus contains a number of horizon-size volumes that varies as $t^{-3/2}$.  Then Eq.~(\ref{t}) shows that the number $N$ of PBHs of mass $M$ formed out of the big volume scales like $\beta M^{-3/2}$. Now the collapse takes about a dynamical time, $GMc^{-3}$, and only thereafter can the process repeat.  Thus $N$ refers to a range of masses $\sim M$ so that the differential distribution is $dN/dM\propto \beta M^{-5/2}$. The power law reflects the lack of a special scale in the inflation spectrum.   The distribution is preserved over time except for dilution by expansion and the dying out due to Hawking evaporation which only leaves PBHs with $M>M_*$.  Accordingly at the present epoch the differential space density is~\cite{carr}
\begin{equation}
dn/dM={\scriptstyle 1\over \scriptstyle 2}M_*{}^{1/2} M^{-5/2}\ \Omega_{\scriptscriptstyle PBH}\, \rho_c\,\Theta(M-M_*)
\label{dn/dM}
\end{equation}
where $\rho_c$ is the present critical density and $\Omega_{\scriptscriptstyle PBH}$ is the fraction of it in PBHs.  We adjusted the numerical factor so that the integral of $M (dn/dM)$ over $M$---the mass density---is precisely $\Omega_{\scriptscriptstyle PBH}\, \rho_c$.

To calculate $\Omega_{\scriptscriptstyle PBH}$ focus, for example, on the end of radiation dominance, $t=t_r,\, z=10^4$, when the horizon volume was $\approx 4(ct_r)^2$.  By today's time $t_0$ that volume has expanded by a factor $(10^{4})^3$.  The present space density of PBHs formed then is obviously $n=\beta\,10^{-12}/(4c^3 t_r{}^3$).   Eq.~(\ref{dn/dM}) gives the alternative estimate $n\approx {\scriptstyle 1\over \scriptstyle 2}M_*{}^{1/2} M_r{}^{-3/2}\ \Omega_{\scriptscriptstyle PBH}\, \rho_c$.  Equating the two, replacing $M_r$ by $c^3 t_r/2G$, and remembering the relation $8G\rho_c\approx H_0{}^2$ between $\rho_c$ and Hubble constant, we get
\begin{equation}
\Omega_{\scriptscriptstyle PBH}\approx 2.5\times 10^{-12}\beta(M_r/M_*)^{1/2} (H_0 t_r)^{-2}= 1.3\times 10^{18}\,\beta.
\label{Om}
\end{equation}
The second equality comes from two observations: for $t>t_r$, $a\propto t^{2/3}$ so that $H_0={\scriptstyle 2\over \scriptstyle 3}t_0^{-1}$; and $t_0/t_r=(10^4)^{3/2}$ (one computes directly $M_r/M_*=6\times 10^{34}$).     Within the scale invariant paradigm (\ref{Om}) is the total PBH density parameter, and Eq.~(\ref{dn/dM}) has predictive power.

Because we must have $\Omega_{\scriptscriptstyle PBH}<1$ we obtain a stiff inequality on $\beta$.  Using Carr's result (based on a Gaussian fluctuation spectrum)
\begin{equation}
\beta\sim \epsilon \exp(-1/18\epsilon^2),
\end{equation}
we get $\epsilon < 0.038$.   This is the accepted bound on primordial fluctuations from inflation~\cite{carr} in the range $M<10^{27}\,\,{\rm Kg}$.  It should be stressed that the scale invariance assumption is a strong one, and even a small $M$ dependence of $\epsilon$ would make $\beta$ strongly mass dependent.  Thus, for example, Carr derives a $M$ dependent version of Eq.~\ref{Om} [Eq.~(7) of Ref.~\cite{carr}], from which he sets constraints on $\epsilon(M)$ separately for each mass range from a variety of astronomical data.  Among these are the known $\gamma$ ray background (which would be affected by just now dying black holes), and the entropy in the CMB which receives a contribution from Hawking radiation.  All in all the search for PBHs makes us much wiser about conditions in the very early universe.

\end{document}